\def\ndla{121}

\def\lya{Ly$\alpha$ }
\def\aap{{A\&A}}

\def\apj{{ApJ}}
\def\apjs{{ApJS}}
\def\apjsupp{ApJS}

\def\mnras{{MNRAS}}
\def\nat{{Nature}}
\def\pasp{{PASP}}
\def\sci#1{{\; \times \; 10^{#1}}}
\def\cm#1{\, {\rm cm^{#1}}}
\def\N#1{{N({\rm #1})}}

\documentclass[cup5b]{caps}
\usepackage{graphicx}
\usepackage{amssymb}
\usepackage{ociwsymp4}   

\HeadText{J. X. Prochaska}

\begin{document}

\pagenumbering{arabic}

\author[]{JASON X. PROCHASKA\\UCO/Lick Observatories}

\chapter{Chemical Abundances in the \\ Damped \lya Systems}

\begin{abstract}

I introduce and review the data and analysis techniques used to measure
abundances in the damped \lya systems, quasar absorption-line systems 
associated with galaxies in the early Universe. 
The observations and issues associated with their abundance analysis
are very similar to those of the Milky Way's interstellar medium. 
We measure gas-phase abundances and are therefore subject to 
the effects of differential depletion.  I review the impact of dust 
depletion and then present a summary of current results on 
the age-metallicity relation derived from damped \lya systems and new 
results impacting theories of nucleosynthesis in the early Universe.

\end{abstract}

\section{Introduction}

While high-resolution stellar spectroscopy provides the framework behind
nearly all discussion of nucleosynthesis and chemical
enrichment (see papers throughout this volume), these observational
efforts are currently limited to the Milky Way and its nearest neighbors.
Beyond the Local Group, it remains very difficult to precisely determine
chemical abundances.  Relative abundance measurements
are limited to a few species (e.g., C, N, O, Ca, Mg) and a few dozen galaxies.
At cosmological distances, the challenges related to galaxy spectroscopy
are even more severe, and even the determination of a crude metallicity
poses great difficulty (e.g., Kobulnicky \& Koo 2000).

Within the Milky Way, absorption-line spectroscopy of the interstellar
medium (ISM) provides abundance measurements for many elements in a range
of physical environments.  
In terms of chemical abundance studies, this analysis is
limited by two factors: (1) gas mixing occurs on short
enough time scales that the majority of the ISM is chemically homogeneous
(Meyer, Jura, \& Cardelli 1998).  Therefore, one cannot probe nucleosynthesis at a range
of metallicity or age in the ISM; and (2) refractory elements like
Si, Ti, Fe, and Ni are depleted from the gas phase.  Their
relative abundance patterns are principally reflective of differential
depletion (see the review by Jenkins 2003).   
ISM absorption-line observations have
also been carried out within the Magellanic Clouds 
(Welty et al.\ 1999, 2001).  These observations reveal the metallicity
of the LMC and SMC, but interpretations of the relative abundance patterns
are complicated by dust depletion. 
Outside of the Milky Way and Magellanic Clouds, it is rarely possible to 
pursue ISM studies in other local galaxies.
With current UV telescopes and instrumentation, there are too few 
bright UV sources behind nearby galaxies.  The advent of the Cosmic Origins
Spectrograph on the {\it Hubble Space Telescope}
will improve the situation, but only to a modest degree.  

Ironically, the laws of atomic physics, the expansion of the
Universe, and the filtering of UV light by the Earth's atmosphere
combine to make the early Universe the most efficient place for 
studying galactic elemental abundances. 
Quasar absorption-line spectroscopy provides a powerful,
accurate means of studying nucleosynthesis and chemical enrichment in
hundreds of galaxies over an epoch spanning several billion
years at $z \approx 2-5$.  These galaxies are called damped \lya systems
(DLAs).
The name derives from the quantum mechanical damping of the \lya profile
owing to the large H~{\sc i} surface density that defines a DLA:
$N$(H~{\sc i}) $\geq\, 2 \sci{20} \cm{-2}$.  At high redshift, the plethora
of UV resonance transitions that ISM researchers study in our Galaxy
are conveniently redshifted to optical wavelengths where they can
be examined using high-resolution spectrographs on 10~m-class
telescopes.  In this fashion, we are able to study the ISM of galaxies
near the edge of the Universe using data that competes with the observations
taken in the Galaxy.  The analysis of the DLAs
has been the focus of our research since the commissioning of the
Keck telescopes, and we are now joined by several groups at observatories
across the globe.

In this paper, I will introduce the techniques used in the analysis
of the DLAs to the broad audience attending the fourth Carnegie Symposium.
By reviewing this topic at a pedagogic level, my goal is to encourage
greater communication between the stellar and 
damped \lya communities.  These
fields of research offer complementary analysis into theories 
of nucleosynthesis and chemical enrichment.  
Although the fields suffer from unique systematic uncertainties, a 
synthesis of their results will ultimately lead to a deeper 
understanding on the origin of the elements.

\begin{figure*}
\centering
\includegraphics[width=1.00\columnwidth,angle=0]{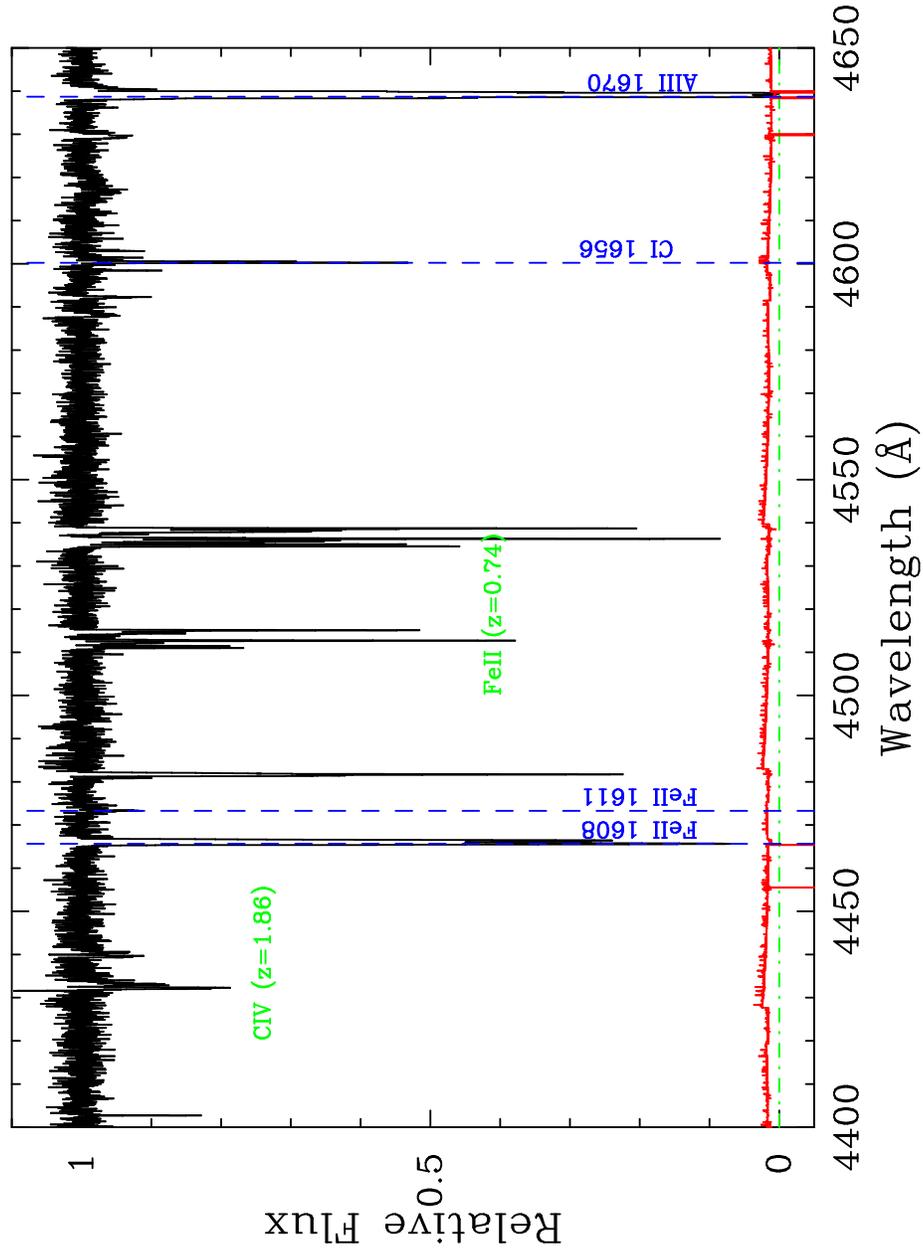}
\vskip 0pt \caption{
A sample HIRES spectrum for the quasar Q1331+17, which exhibits 
a $z=1.776$ DLA whose transitions are marked by vertical dashed lines.
The S/N of these data is somewhat higher than most observations, while the
resolution is typical (FWHM~$\approx 7$~km~s$^{-1}$).   In addition to the 
DLA transitions, there are several absorption lines arising from 
``metal-line systems'' at $z=0.74$ and $z=1.86$.}
\label{fig:data}
\end{figure*}

\section{The Data and Standard Analysis}

The data that drive current elemental abundance research in the DLAs
is very similar to the spectroscopy obtained for
stars in the Milky Way.  The largest data sets to date have been
acquired with the HIRES
echelle spectrograph (Vogt et al.\ 1994) on the Keck~I telescope and
the UVES-VLT spectrograph (Dekker et al.\ 2000).  These spectrographs
do provide resolution $R \geq 60,000$, yet the majority of damped \lya
research is conducted with $R \approx 40,000$ observations.  
While the ``clouds'' of gas that comprise the velocity profiles of the
galaxies presumably have thermal widths below the resolution of the
spectrograph, tests for ``hidden'' saturation (e.g., Prochaska \& Wolfe 1996)
have demonstrated that line saturation is not an important issue.
Even the lower-resolution DLA studies by Pettini et al.\ (1994, 1997) on
4~m-class telescopes gave rather accurate column density measurements for
the weak, unsaturated Zn~{\sc ii} and Cr~{\sc ii} profiles.  

Figure~\ref{fig:data}
presents a sample of data showing several metal-line transitions
for a typical damped \lya system.  
The dashed vertical lines identify four transitions related to
the $z=1.776$ DLA, and we also identify a C~{\sc iv} doublet and Fe~{\sc ii}
multiplet related to two other absorption systems along the sightline.
The line density described by this
figure is typical of quasar spectra redward of the \lya forest,
i.e.\ $\lambda > (1+z_{\rm QSO}) \times 1215.67$ \AA.  Clearly, 
line blending is a rare phenomenon and spectral synthesis is
generally unrequired.  
This is contrasted, of course, by analysis within the \lya forest
where contamination by coincident \lya clouds is common.
Another point to emphasize is that the majority of absorption 
systems along a quasar sightline tend to show absorption from C~{\sc iv},
Si~{\sc iv}, and Mg~{\sc ii} doublets.  These are trivially identified, and 
therefore line misidentifications are very unlikely in the analysis
of the damped \lya systems.

For many transitions, one approaches 1\% statistical error in the column
density measurements with a signal-to-noise ratio (S/N) per pixel of only 30.  
Therefore, few (if any) of the damped \lya systems have been observed
at S/N~$> 100$ per pixel or even the S/N of the data in Figure~\ref{fig:data}.
Column densities are generally determined from either the summation 
of the observed optical depth (Savage \& Sembach 1991) or through
detailed Voigt-profile fitting.  Oscillator strengths for the dominant 
transitions are almost exclusively from laboratory measurements and have
typical errors of $< 20\%$.  
There are several important exceptions, however, notably
the Fe~{\sc ii} $\lambda$1611 transition (one of the key Fe~{\sc ii} transitions
for metal-rich DLAs)  whose best value is based on
a theoretical calculation.  
Regarding the relative oscillator strengths, most
of these transitions have been
extensively analyzed in the Galactic ISM, and inaccuracies in the oscillator
strengths have been corrected in the literature (e.g., Zsarg$\rm \acute o$ \&
Federman 1998; Howk et al.\ 2000).

\begin{figure*}
\centering
\includegraphics[width=0.75\columnwidth,angle=-90]{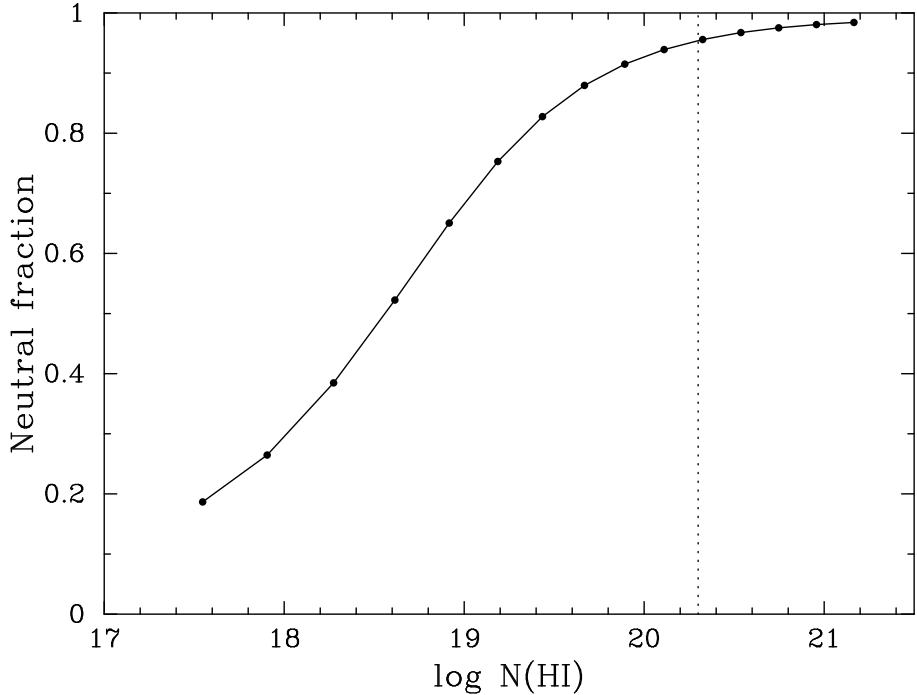}
\vskip 0pt \caption{
Predicted neutral fractions for a plane-parallel slab of hydrogen
gas with a range of H~{\sc i} surface densities $N$(H~{\sc i}).  In this
calculation, we assumed a hydrogen volume density $n_{\rm H} = 0.1 \cm{-3}$
and a standard extragalactic background radiation field.  The results are
based on a standard radiative transfer calculation (see Prochaska \& Wolfe
1996 for more details).
}
\label{fig:ion}
\end{figure*}

Perhaps the most startling aspect of damped \lya research for stellar
spectroscopists is that the path from ionic column density measurements
to elemental abundances is trivial.  One often observes only
one ion per element, the dominant species in a neutral hydrogen gas,
and assumes no ionization corrections to compute gas-phase 
elemental abundances.  The neglect of ionization corrections was unavoidable
in the past (the first sets of observations
provided few, if any, diagnostics) and was supported by theoretical
expectations (Viegas 1994; Prochaska \& Wolfe 1996) as well as
observations of H~{\sc i} clouds in the Milky Way ISM with column densities
comparable to the DLAs (see Jenkins 2003).  
Figure~\ref{fig:ion} shows a simple radiative transfer calculation
for an ionizing flux incident on a plane-parallel, constant-density 
slab of hydrogen gas (Prochaska \& Wolfe 1996).
For an assumed number density of $n = 0.1 \cm{-3}$, 
we find that a damped \lya system
with $N$(H~{\sc i}) $>\, 10^{20.3} \cm{-2}$ would be $<  10\%$ ionized.  
More accurate and realistic
calculations have come to similar conclusions (Vladilo et al.\ 2001),
yet empirical confirmation remains an outstanding problem.  
Prochaska et al.\ (2002a) presented one of the few cases (the DLA at $z=2.625$
toward Q1759+75) where transitions
from multiple ionization states were unambiguously detected and argued that this 

DLA with $N$(H~{\sc i}) $=\, 10^{20.65} \cm{-2}$
requires significant ionization corrections, contrary to the 
theoretical expectation.  These authors also argued, however, that 
the conditions in the DLA toward Q1759+59 are probably unusual.  Indeed,
this expectation is supported by more
recent studies (e.g., Prochaska et al.\ 2002b).

Without ionization corrections, the gas-phase abundances are trivially
computed from the ionic column densities (e.g., Fe/H = $\N{Fe^+}/\N{H^0}$),
and the precision of the elemental abundances match those of the 
column densities.  Owing to systematic errors related to continuum fitting
near the \lya profile, the principal source of error tends to lie in
$\N{H^0}$, which generally limits the precision to $\sim 0.1$~dex.  
For relative abundance measurements the precision is often
better than 0.05~dex at the $1 \sigma$ level, surpassing all but the most
accurate relative abundance measurements derived from stellar analyses.
As we shall see in the next section, however, the precision achieved for
these gas-phase abundances can be severely compromised by the effects of dust.

\section{Dust}

Since the pioneering studies by Str\"omgen (1948) and Spitzer (1954) on Ca$^+$ 
and Na$^0$ ions in the Milky Way ISM, astronomers have appreciated 
that refractory
elements are depleted from the gas phase.  These optical surveys gave 
the first convincing demonstration of depletion, and later 
UV spectroscopy revealed a more
complete picture of dust.  Similar observations of gas within
the SMC and LMC have provided insight into depletion in other galactic
systems and have suggested depletion is universal with a generic pattern
(see the Jenkins 2003 review).

Presently, the majority of the uncertainty, confusion, frustration, and pain
associated with studying chemical abundances in the damped \lya systems
stems from dust (see the reviews by Draine 2003 and Jenkins 2003 for a complete
discussion of dust).  Dust plays two roles in the observations, one direct
and one indirect.  The direct effect is that refractory elements in the
DLA (e.g., Fe, Ni, Cr) are depleted from the gas phase into and onto
dust grains.  One expects
the processes are similar to those observed for the ISM of the Milky
Way, although it is difficult to confirm this at high redshift.  It is
clear, however, that 
depletion levels in the DLA are significantly
lower than typical sightlines through the Milky Way, instead often 
resembling warm gas in our Galactic halo or the gas in the LMC and SMC.

Dust would not pose such a difficult
problem in a discussion of the DLA abundances if not for two
points: (1) spectra of the ``typical'' DLA generally allow abundance
measurements for only a few elements, primarily Si, Fe, Ni, Cr, and Zn; 
(2) there is an unfortunate degeneracy
between the differential depletion patterns of these few
elements (Zn exempted)
and the nucleosynthetic pattern expected for Type~II supernovae
(e.g., Woosley \& Weaver 1995).  To wit, differential depletion implies
enhancements of Si/Fe and roughly solar Fe, Cr, and Ni abundances, as does
Type~II supernova nucleosynthesis\footnote{I suspect that this degeneracy between 
seemingly very different processes may not be a simple coincidence
but may be the result of condensation temperatures correlating with 
even-numbered nuclei.}. It is for this reason, above all others, that the 
non-refractory element Zn, an element with a speculative nucleosynthetic 
origin {\it at best}, has received such great prominence in DLA abundance
research.
Empirically, Zn roughly traces Fe in stars with metallicity [Fe/H]~$> -3$
(see the review by Nissen 2003), and Zn is very nearly non-refractory.
Therefore, the majority of the DLA community has adopted Zn as a proxy for
Fe and have imposed dust corrections on the gas-phase abundances under
the assumption that Zn/Fe should be solar in the DLAs (e.g., Vladilo 1998).
These are sensible approaches, but the uncertainty in the
nucleosynthetic origin of Zn gives me pause
(as do issues relating to ionization corrections; see Jenkins 2003).
If the Zn/Fe ratio is intrinsically $\sim +0.2$~dex
in the DLA, one may draw very different conclusions on the
$\alpha$/Fe ratios and ultimately the roles of various supernovae in the
enrichment of these galaxies. 

The other major issue related to dust in the DLAs is obscuration.
The vast majority of DLAs have been identified toward bright quasars identified
in optical or UV surveys.  If the sightline to a given quasar penetrates a 
gas ``cloud'' with a large column of dust, it is possible the quasar will
be removed from these magnitude-limited surveys.  This was the concern of 
Ostriker \& Heisler (1984), and Fall \& Pei (1993) have developed a formalism
to account for this selection bias.  A full discussion of the likelihood
that dust obscuration is influencing studies of damped \lya abundances
is somewhat beyond the scope of this paper.  We will return to the topic
in the next section, but we also refer the interested reader to the 
papers by Boiss\'e et al.\ (1998), Ellison et al.\ (2001), and Prochaska \& 
Wolfe (2002).
My hope and current expectation is that the effects of dust
obscuration are small at high redshift where the gas metallicities are lower
and the observed sightlines show relatively low depletion levels and low
molecular gas fractions (Ledoux, Petitjean, \& Srianand 2003).

\section{Chemical Evolution}

The zeroth-order measure of a damped \lya system (aside from its redshift)
is the H~{\sc i} column density.  By surveying the \lya profile toward quasars
at a range of redshifts, observers have traced the H~{\sc i} mass density of the
Universe at a range of epochs (Wolfe et al.\ 1986; Lanzetta, Wolfe, \& 
Turnshek 1995; Wolfe et al.\ 1995; Storrie-Lombardi \& Wolfe 2000).
The measurements for the damped systems provide a cosmic H~{\sc i} mass
density because these galaxies dominate the neutral hydrogen gas
density to at least $z=4$ (see also P${\rm \acute e}$roux et al. 2003).

The first-order measure of a DLA is its metallicity.  This is
determined from the gas-phase measurements
of Fe$^+$, Si$^+$, Zn$^+$, Cr$^+$, and other ions.  Because of dust depletion,
one expects that the refractory elements (e.g., Fe, Ni, Cr) provide
systematically lower metallicity values.  When possible, therefore, 
observers have focused on non-refractory or mildly refractory elements,
especially Zn.  Indeed, this dictated the strategy of the first surveys
by Pettini et al.\ (1994, 1997).  These surveys provided the
first $\sim 20$ DLA metallicities, which showed the mean metallicity of the
DLA (i.e.\ the neutral gas of the Universe)  is $\sim 1/10$ solar at $z=2$.

In the 10 years since Pettini and collaborators initiated this field,
the study of DLA abundances has evolved substantially,
primarily owing to the birth of 10~m-class telescopes.
Surveys with HIRES on the Keck Telescope (Lu et al.\ 1996; 
Prochaska \& Wolfe 1999, 2000) have pushed the metallicity measurements to
$z=4$ and beyond and allowed the first examination of evolution in the
mean metallicity.  To extend the metallicity measurements above $z=3$,
these authors had to consider elements other than Zn because (1) it is
difficult to measure its weak transitions along low-$N$(H~{\sc i}), 
low-metallicity sightlines and (2) the Zn~{\sc ii} transitions are redshifted 
to observed wavelength
$\lambda > 9000$ \AA.  The Prochaska \& Wolfe (2000) survey was comprised
of $\sim 50$ Fe measurements ranging from $z=2$ to 4 and showed no
statistically significant evolution in the mean metallicity.  The advent
of UVES on the VLT has led to an additional set of measurements
(e.g., Molaro et al.\ 2000; Dessauges-Zavadsky et al.\ 2001), and the
largest recent impact comes from new surveys using the
Echellette Spectrograph and Imager (ESI) on the Keck~II Telescope
(Prochaska et al.\ 2003a, c).  Owing to the high
throughput of this moderate resolution spectrograph ($R \approx 10,000$)
and an improved observing strategy, we have roughly doubled the sample of $z>2$
metallicity measurements in $\sim 1/10$ the observing time.
For the foreseeable future, instruments like ESI are going to lead this area
of damped \lya research.

\begin{figure*}
\centering
\includegraphics[width=0.75\columnwidth,angle=-90]{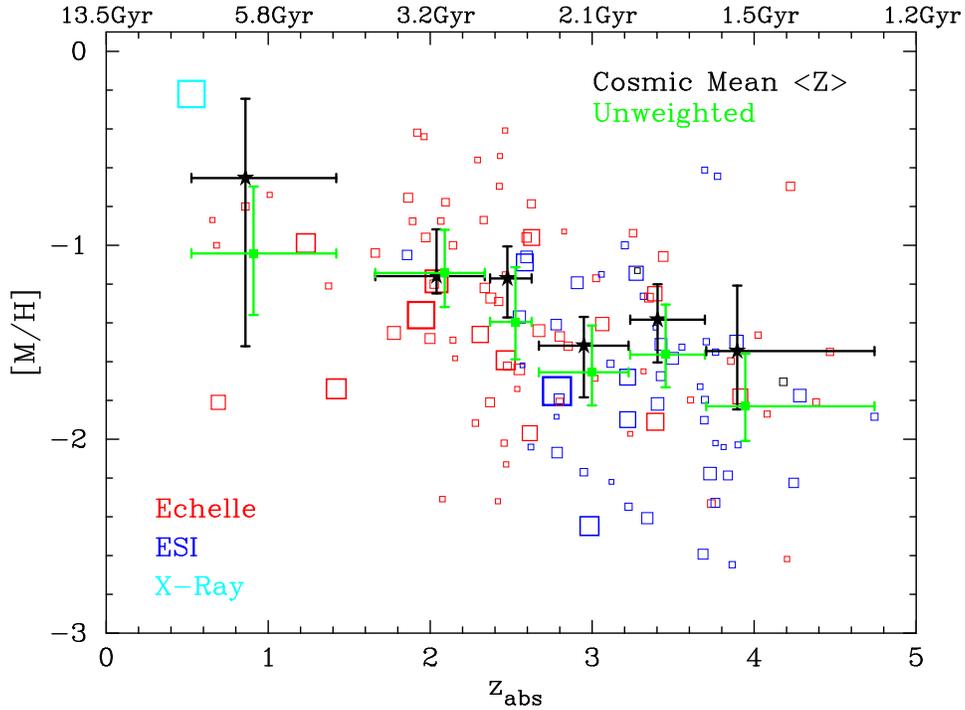}
\vskip 0pt \caption{
Summary of the metallicity measurements vs.\ redshift for the \ndla\
DLAs comprising the full, current sample.  The area of the data points (squares)
scales with the $N$(H~{\sc i}) values of the DLAs.  The dark binned values with
stars correspond to the cosmic mean metallicity $\langle Z\rangle$, which is
the metallicity of the Universe in neutral gas.}
\label{fig:chemE}
\end{figure*}

Figure~\ref{fig:chemE} presents a summary of the current set of damped
\lya metallicities, [M/H], as a function of redshift.  
A detailed discussion of these results is given in
Prochaska et al.\ (2003b).  In brief, both the unweighted and H~{\sc i}-weighted
mean metallicities show evolution with redshift at $3 \sigma$
significance with a slope $m \approx -0.25$~dex/$\Delta z$.
The H~{\sc i}-weighted mean, often denoted $\langle Z\rangle$, is a true cosmic quantity; it
represents the mean metallicity of the Universe in neutral gas.
Its determination allows direct comparisons with chemical enrichment
and galaxy formation models in the early Universe (Pei, Fall, \& Hauser 1999;
Mathlin et al.\ 2000; Somerville, Primack, \& Faber 2001).
At present, there is a significant disagreement between the metallicities
implied by the DLAs and the metal production inferred from a derivation
of their star formation rates via observations of the C~{\sc ii}$^*$ $\lambda$1335 transition.
(Wolfe, Gawiser, \& Prochaska 2003).  This
``missing metals'' problem raises an important challenge for future
observations and theoretical efforts related to the production of metals
in the early Universe.

\begin{figure*}
\centering
\includegraphics[width=0.75\columnwidth,angle=-90]{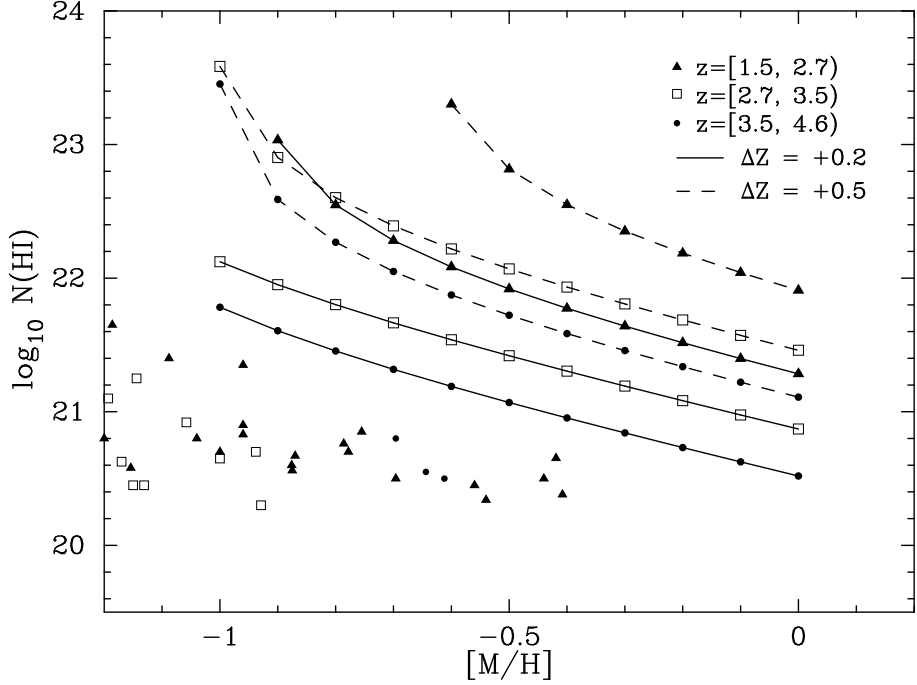}
\vskip 0pt \caption{
This figure describes the robustness of the $\langle Z\rangle$ values presented
in Fig.~\ref{fig:chemE} to the presence of an outlier.  We characterize
the outlier by a range of H~{\sc i} column densities and [M/H] values.  The
curves are contours of constant $\Delta Z$, the change in $\langle Z\rangle$ from including
an outlier as a function of $N$(H~{\sc i}) and [M/H].  The point styles refer
to three redshift intervals.  The ``free'' points in the figure are observed
DLA galaxies.}
\label{fig:robust}
\end{figure*}

\begin{figure*}
\centering
\includegraphics[width=0.75\columnwidth,angle=-90]{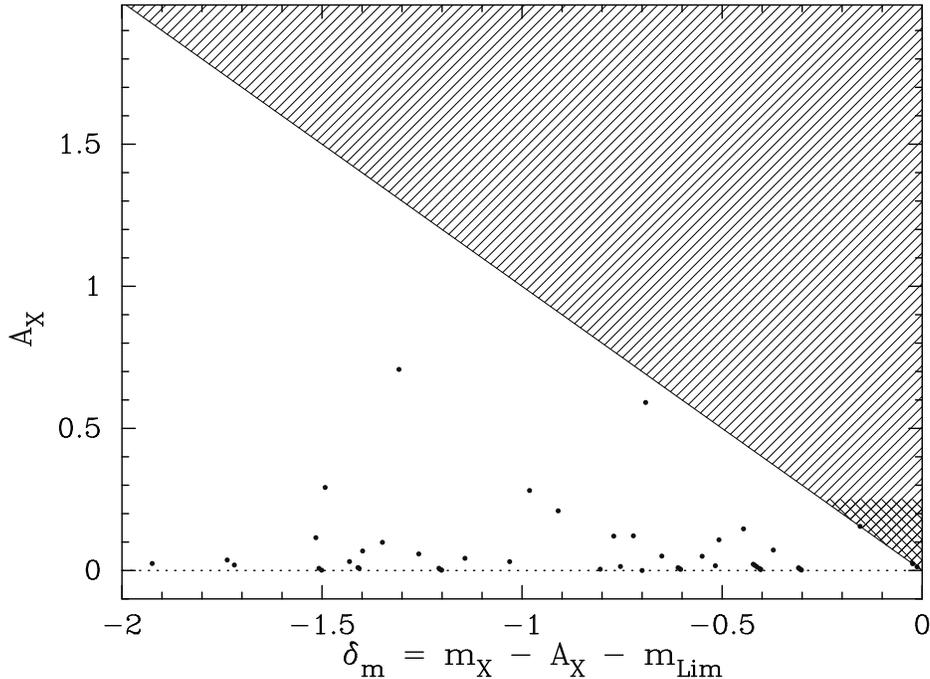}
\vskip 0pt \caption{
Extinction corrections $A(X)$ derived from observed dust
column densities
against $\delta_m$, the difference between the corrected brightness
of the background quasar relative to the limiting magnitude of
the damped \lya survey (see Fig.~23 of Prochaska \& Wolfe 2002).  
The shaded region denotes
the area of parameter space that obscured quasars would occupy.  
The cross-hatched region designates the area of parameter space
that we contend is populated by obscured quasars with foreground
damped \lya systems
as inferred by the observed distribution of $A(X), \delta_m$ values.
}
\label{fig:Amag}
\end{figure*}

Prochaska et al.\ (2003c) noted that the set of \ndla~damped systems is
the first sample with sufficient size to present a determination
of $\langle Z\rangle$ robust to ``reasonable'' outliers.
This point is emphasized in Figure~\ref{fig:robust}, where we plot contours
of $\Delta Z$, the change in $\langle Z\rangle$, as a function of 
$\log$$N$(H~{\sc i}) and [M/H]
values for an assumed outlier.  The point types correspond to
various redshift bins and the line style indicates the magnitude change
in $\langle Z\rangle$.  Pairs of $\log$$N$(H~{\sc i}), [M/H] values for the observed DLA 
are also shown in the figure
as isolated points with point type according to their redshift.
Consider the interval $1.5<z<2.7$.  To impart a 0.2~dex change
in $\langle Z\rangle$, one requires an outlier with 1/3 solar metallicity to have
$N$(H~{\sc i}) $\approx\, 10^{22} \cm{-2}$.
An outlier with these characteristics would
lie one magnitude off the observed distribution of $N$(H~{\sc i}), [M/H] values.  Even
in the highest-redshift interval, which has the smallest sample size, 
it would take an outlier with 
a product of $N$(H~{\sc i}) and metallicity that is 3 times larger than any
current observation.
By definition, of course, an outlier lies off the main 
distribution of observed values.
At present, however, to impose a large increase in $\langle Z\rangle$, one would have 
to introduce an outlier that lies far beyond the distribution of 
observed values.  If such an outlier is identified,
it would strongly suggest the existence
of a currently unidentified population of DLAs with 
$\log$$N$(H~{\sc i}) + [M/H] $>$ 20.6.

\begin{figure*}
\centering
\includegraphics[width=0.75\columnwidth,angle=-90]{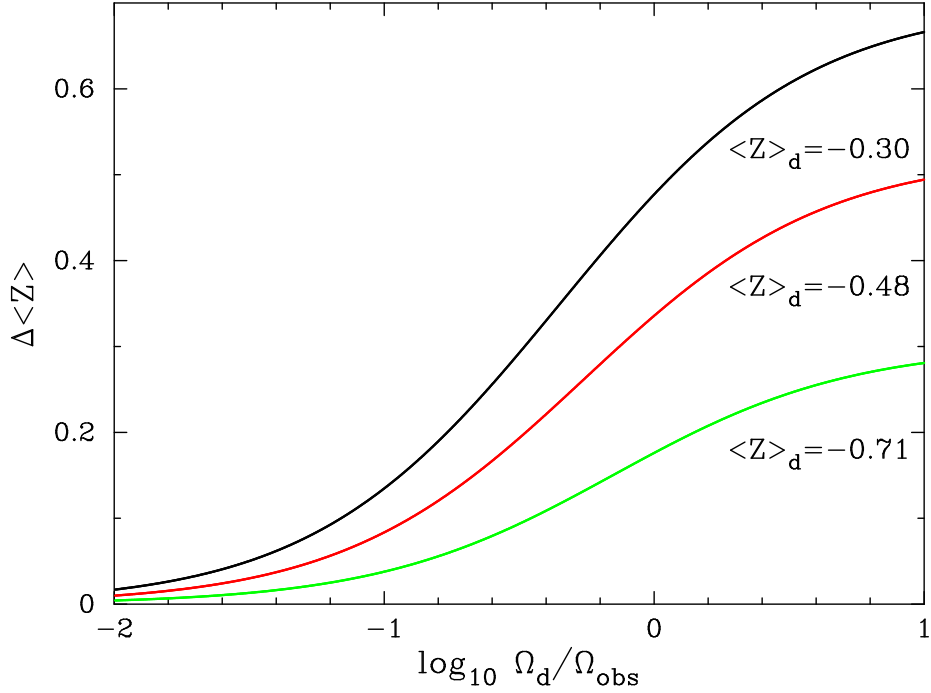}
\vskip 0pt \caption{
Correction ($\Delta Z$) to a $\langle Z\rangle$ value of 1/10 solar
for an obscured gas component with mass density $\Omega_{\rm d}$ and average
metallicity $\langle Z\rangle_{\rm d}$.  The figure shows that if the observed 
gas mass density equals the observe quantity ($\Omega_{\rm d}/\Omega_{\rm obs} 
= 1$) and the observed gas has metallicity $\langle Z\rangle$ = 1/3 solar, it 
would imply a factor of 2 correction to the observed mean metallicity.
}
\label{fig:obsc}
\end{figure*}

Indeed, Boiss\'e et al.\ (1998) were the first to emphasize that the observed
DLAs exhibit an upper limit to the sum $\log$$N$(H~{\sc i}) + [M/H].  They
interpreted this upper limit in terms of dust obscuration; DLAs with a large
product of $N$(H~{\sc i}) and metallicity may have larger dust optical depths
and therefore may significantly obscure background quasars at UV wavelengths.
We have argued that no evidence exists
for significant obscuration at $z>2$, as Ellison et al.\ (2001) and 
Prochaska \& Wolfe (2002) discuss.   
The former authors have conducted a survey of DLAs toward radio-selected quasars
and found no significant difference in the H~{\sc i} content of these DLAs, contrary
to the expectation for dust obscuration.
The latter authors have argued that the observed distribution of inferred
extinction values
and apparent magnitudes indicate depletion plays a minor effect in the
DLA analysis.  In Figure~\ref{fig:Amag}, we present a figure similar to
Figure~23 of Prochaska \& Wolfe (2002), which includes the majority of the DLAs
from our ESI survey.  We refer the reader to the discussion in Prochaska \& Wolfe
(2002) for further details.

Independent of the above arguments, it is possible to assess
the effects that dust obscuration will have on the measurements of $\langle Z\rangle$ for
an assumed mass and metallicity of the obscured gas. 
Let $\langle Z\rangle_{\rm tru} \, = \, \langle Z\rangle + \Delta Z$ and 
$\Omega_{\rm tru} = \Omega_{\rm obs} + \Omega_{\rm d}$, where $\Omega_{\rm obs}$ is the 
observed neutral gas density and $\Omega_{\rm d}$ is the gas density that 
is obscured.  Finally, assume that the obscured
gas has mean metallicity $\langle Z\rangle_{\rm d}$.  Figure~\ref{fig:obsc} shows the
correction $\Delta Z$ to an observed mean metallicity of $\langle Z\rangle$ = 
 --1~dex for
a range of $\Omega_{\rm d}/\Omega_{\rm obs}$ values and three assumed $\langle Z\rangle_{\rm d}$ values.
We emphasize that the results in Figure~\ref{fig:obsc} hold independently 
of the number of DLAs observed in the determination of $\langle Z\rangle$.
The CORALS survey (Ellison et al.\ 2001) has argued that 
$\Omega_{\rm tru} \approx \Omega_{\rm obs}$, implying that $\Omega_{\rm d} / \Omega_{\rm obs}$ 
is small.  If $\Omega_{\rm d} / \Omega_{\rm obs}$ is as large as 1, then one must
worry about a significant correction to $\langle Z\rangle$ for 
$\langle Z\rangle_{\rm d}$ values greater than
1/3 solar.  If future surveys similar to CORALS demonstrate that 
$\Omega_{\rm d} / \Omega_{\rm obs}$ is 1/10 or smaller, then it is unlikely dust 
obscuration will ever play a major role in the determination of $\langle Z\rangle$.

\section{Nucleosynthesis}

While echelle observations on 10~m-class telescopes have led to a greater
number of DLA metallicity measurements at a greater range of redshifts, 
the most significant advances from large telescopes
have come through
studies of the relative chemical abundances.  These observations
reveal the processes of nucleosynthesis during the first few
billion years of the Universe and ultimately provide insight into the nature
of the DLA galaxies and scenarios of galaxy formation (see 
Calura et al. 2003).

As discussed above, typical DLA observations have yielded relative abundance
measurements for Si, Fe, Cr, Ni, and Zn with accuracies better than 10$\%$.
The principal difficulty in applying these observations to studies of 
nucleosynthesis is the effects of differential depletion.  Specifically, the
standard ISM depletion patterns (Savage \& Sembach 1996)
are very similar to the patterns expected
for the yields from Type~II supernovae (e.g., Woosely \& Weaver 1995).  
This degeneracy led to 
arguments over the appropriate interpretation of Si/Fe enhancements and
other ratios resulting from the first DLA surveys
(Lu et al.\ 1996; Vladilo 1998; Pettini et al.\ 1999;
Prochaska \& Wolfe 1999).  These
arguments are still being discussed and may never be unambiguously resolved.
Therefore, the community has turned its attention toward obtaining observations
of elements that are not heavily depleted (e.g., N, O, Si, S, Zn, P) or whose
differential depletion pattern runs contrary to expectations from 
nucleosynthesis (e.g., Mn/Fe, Ti/Fe; Dessauges-Zavadsky, 
Prochaska, \& D'Odorico 2002).

\begin{figure*}
\centering
\includegraphics[width=0.75\columnwidth,angle=-90]{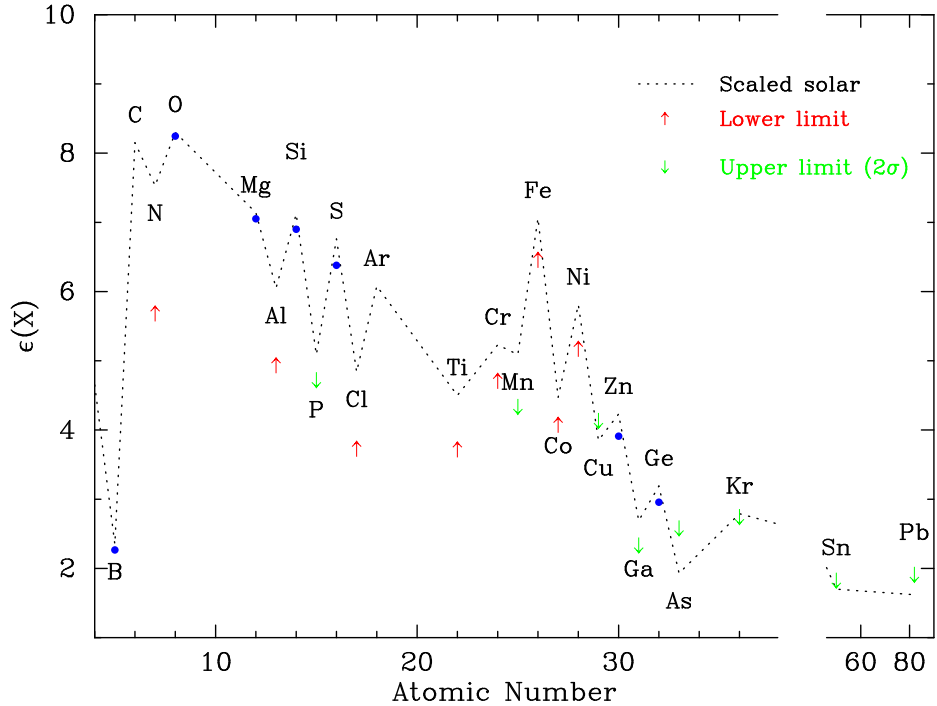}
\vskip 0pt \caption{
Abundance pattern of the $z=2.626$ DLA toward FJ0812+32.
The solid circles and arrows represent detections and limits, respectively.
The dotted line shows the solar abundances scaled to the oxygen abundance
in the DLA.}
\label{fig:fj0812}
\end{figure*}

One example of a nucleosynthetic diagnostic that is nearly free of depletion
effects is the N/$\alpha$ ratio (where $\alpha$ in the DLA is generally
given by Si and S).  This abundance ratio is an excellent 
diagnostic of the time scales of star formation owing to the 
belief that N production is dominated by intermediate-mass stars (see
Henry 2003).
For this reason, among others, observations of N/$\alpha$ have played an 
important role in several contributions to this Symposium (see papers by, e.g., Garnett 2003 and Molaro et al. 2003). 
Regarding the DLAs over the past two years,
observers have built a sample of $\sim 30$ measurements with metallicities
ranging from [$\alpha$/H]~=~$-2$ to nearly solar (Pettini et al.\ 2002;
Prochaska et al.\ 2002b; Centuri\'on et al.\ 2003).  
In general, these measurements track the N/$\alpha$ values observed in 
H~{\sc ii} regions and stars in the local Universe.
With our sample (Prochaska et al.\ 2002b), however, we speculated
that the DLAs at low metallicity show 
a bimodal distribution of N/$\alpha$ values.  In
particular, we found that a small but significant fraction of these DLAs have very low 
N/$\alpha$ values, much lower than any value observed locally.
If confirmed by future surveys, this bimodality may
require revised yields of N in massive stars (e.g., Molaro et al. 2003), an 
initial (Population III?)
epoch of star formation characterized by a top-heavy or truncated initial mass function 
(Prochaska et al.\ 2002b), or some other unappreciated
physical mechanism.  We defer additional discussion of this topic to the
paper by Molaro et al. (2003).

At this Symposium, we reported the discovery of a DLA whose large $N$(H~{\sc i})
and [M/H] values will allow the detection and analysis of over 20 elements
in a single galaxy (Prochaska, Howk, \& Wolfe 2003).  Figure~\ref{fig:fj0812}
shows the abundance pattern of this galaxy and a comparison with the solar
abundance pattern scaled to the galaxy's oxygen abundance ([O/H] $\approx$ 1/3 solar).
Aside from the analysis of stars in the Milky Way and its nearest neighbors
(see, e.g.,  Hill 2003; Shetrone 2003; Venn et al. 2003), galaxies like this DLA 
will enable the most comprehensive nucleosynthetic analysis at any epoch in the
Universe.  The results presented in 
Figure~\ref{fig:fj0812} and future observations will have the following
impacts on theories of nucleosynthesis.

\begin{enumerate}
\item
Observations of the B/O ratio test processes
of cosmic ray and $\nu$-wind spallation invoked to explain the
production of the light elements B, Be, and Li (Fields \& Olive 1999).

\item
Some of the galaxies will show measurements of all three CNO 
elements in the same DLA.  These measurements yield 
clues to nucleosynthesis in intermediate-mass stars
and place important time constraints on metal enrichment 
(Henry, Edmunds, \& K{\"o}ppen 2000).

\item
The relative abundances of the $\alpha$-elements (e.g., O, Mg, Si, S)
and the examination of odd-Z elements (e.g., P, Al, Ga, Mn)
can be used to test predictions of explosive nucleosynthesis 
(e.g., Woosley \& Weaver 1995).  Furthermore, these abundances probe the 
initial mass function and mix of Type\,II vs.\ Type\,Ia supernovae in these 
protogalaxies.  For the galaxy in Figure~\ref{fig:fj0812}, the
decline in relative abundance of the $\alpha$-elements (e.g., [O/S]~$\approx +0.3$)
and the enhanced ``odd-even effect'' (e.g., [P/Si]~$< 0$) suggest
an enrichment history dominated by massive stars.

\item
Observations of Pb, Kr, Sn, and Ge will constrain theories of
the $s$-process and $r$-process, and particularly AGB nucleosynthesis 
(e.g., Travaglio et al.\ 2001).
It is important to emphasize that this scientific inquiry takes place
in a relatively metal-rich gas (O/H~$\approx 1/3$~solar) in a system
with a strict upper limit to its age of 2.5~Gyr.
The latter point is particularly relevant to theories on nucleosynthesis
because this time scale limits the contribution from intermediate-mass stars.
\end{enumerate}

While the results for the galaxy
presented in Figure~\ref{fig:fj0812} will---on their own---place new 
constraints on theories of nucleosynthesis, the real 
excitement from its discovery is the promise of identifying
many other galaxies with similar characteristics.
We are currently pursuing several DLAs with $N$(H~{\sc i}) and [M/H]
values similar to those of the $z=2.626$ DLA toward FJ0812+32, whose
observations should yield abundance measurements of $\sim 20$
elements in each DLA.  These observations will reveal if the $z=2.626$
DLA toward FJ0812+32 is a unique case or representative of the population
of metal-rich DLAs.
In addition to these efforts, we have begun a survey with ESI on 
Keck~II to find an additional 10--50 of these DLAs.
Several groups at this meeting are now involved with searches for
extremely metal-poor stars at $z=0$.  Our complementary effort is to
discover relatively metal-rich galaxies at very high $z$.

\end{document}